# NOISE AND ARTIFACTS ELIMINATION IN ECG SIGNALS USING WAVELET, VARIATIONAL MODE DECOMPOSITION AND NONLOCAL MEANS ALGORITHM

Taoufik Ben Jabeur [1*][iD], Eihab Bashier[2][iD], Qudsia Sandhu[iD][1], Kelvin Joseph Bwalya[iD][2], Adason Joshua[iD][3]

[1] Dhofar University, P.O.Box:2509 | Postal Code: 211 | Salalah, Sultanate of Oman.
[2] Sohar University, P.O Box 44, Al Jameah Street, Sohar, Al-Batinah, Oman
[3] Lifeline hospital, Hamdan Building, Al Wadi, Salalah, Sultanate of Oman. P.O.Box No: 190, P.C: 211

Corresponding Author Email: tjabeur@du.edu.om

## ABSTRACT

Electrocardiogram (ECG) signals can frequently be affected by the introduction of noise and artifacts. Since these types of signal corruptions disrupt the accurate interpretation of ECG signals, noise and artifacts must be eliminated during the preprocessing phase. In this paper, we introduced a comprehensive pre-processing phase that eliminates motion artifacts and noise prior to detecting and extracting entirely corrupted ECG signal segments. The first method, denoted as the WLNH method, is constructed using Wavelet multiresolution analysis (MRA), the Lillifors test, NLM, and a high pass filter. The second method entails substituting the Wavelet MRA decomposition with the Variational Mode decomposition (VMD), while retaining all other stages from the first method. This technique is denoted as the VLWNH. The two proposed methods differ from some existing methods in that they first employ the Lilliefors test to identify whether a component is white Gaussian noise and then utilize the High Pass Filter to eliminate motion anomalies. The simulation results show that the offered solutions are effective, particularly when dealing with white gaussian noise and base-line wander (BW) noise.

## I. Introduction

Cardiovascular disease (CVD) can occur when there is any malfunction in the cardiovascular system, which is responsible for pumping blood throughout the circulatory system [1, 2]. In 2019, 17.9 million CVD-related fatalities occurred worldwide, accounting for 32% of total deaths [3].
According to statistics, coronary artery disease (CAD) was the leading cause of global mortality in 2019. It was responsible for around 16.6% of all fatalities. This indicates that CAD causes 27,624 deaths each day throughout the world [4].
Multiple studies and efforts in the literature have concentrated on early detection of cardiovascular diseases (CVDs) with the aim of decreasing mortality rates. There are several diagnostic procedures for cardiovascular disease (CVD) amongst which, three stand out: the blood sample method, cardiac magnetic resonance imaging, and echocardiography.
The advantages and limitations of each approach are outlined below:

1) **Blood sample Method:** Utilizing a conventional blood test can facilitate the identification and assessment of certain diseases and disorders. The use of the protein "troponin" in a blood test provides a rapid and precise assessment of any heart muscle damage [5]. Nevertheless, the analysis of blood necessitates costly apparatus and the constant presence of a nurse or healthcare practitioner. Furthermore, blood tests have a limited duration of usefulness.

2) **Cardiac Magnetic Resonance Imaging method (CMRI):** The cardiac magnetic resonance imaging (CMRT) approach has several important advantages, as it allows accurate visualization of the function of the various components of the heart, such as the ventricles, atria, valves, muscles, and blood flow [6]. Unfortunately, CMRT is cost-prohibitive and is also an uncomfortable test. Additionally, it can be invasive to use for long periods of time.

3) **Echocardiography Method:** The electrocardiogram, abbreviated as EKG or ECG, examines the electrical activity of the heartbeat by analyzing the recorded signal provided by the electrical activities [7]. This procedure is straightforward, safe, and inexpensive. Furthermore, ECG signals may be recorded using a variety of instruments, including medical equipment and smart watches, and they can be seen and interpreted in real time [8]. However, a clinician has to interpret the recorded ECG data.

Out of the three possible test procedures, only the final one is applicable for achieving the purpose of continually monitoring the patient's heartbeat.
Nowadays, we are witnessing rapid technological advancements in both hardware and software. Therefore, it is possible to find relatively small devices to record ECG signals using surface electrodes in hardware. Unfortunately, most of the recorded ECG signals are corrupted by noise such as baseline wander (BW), muscle artifact (MA) as a result of periods of muscle contraction, instrumentation noise (IN), and powerline interference (PLI) from electrical activity, as well as artifacts.
In order to interpret the ECG signals correctly, a preprocessing stage is required to eliminate the noise and artifacts. Several techniques appeared in the literature for denoising ECG signals, including low-pass filters [9], wavelets, empirical mode decomposition (EMD), least mean squares (LMS), Deep learning, etc. Most of these techniques are reviewed and detailed in [10]. Recently, a new technique of signal denoising is proposed in [11] based on the use of EMD and VMD jointly. Furthermore, some studies also focused on motion artifact elimination. In [12], an adaptive algorithm based on Recursive Learning Square (RLS) and low pass filter is proposed to eliminate the motion artifacts. By using VMD and Discrete Wavelet transform, a new method of motion artifact elimination is proposed in [13].
To the best of our knowledge, all previous studies focused either on noise elimination or motion artifacts elimination. Few works were proposed to eliminate them jointly.

There are several data sets of ECG signals in the literature, mostly can be found in the MIT-BIH Arrhythmia database. This research considers both the MIT-BIH Arrhythmia database [14] and the INCART public dataset from the St. Petersburg Institute of Cardiological Technics (Russia) [15].

This paper can be considered as an extension of the recent work in [13]. We exploit the idea of the presence of two categories of components located in low and high pass bands. After signal decomposition via wavelet MRA or VMD, we introduce in this paper the use of Lilliefors test in the first category of components located in a high frequency band to find out any component as white gaussian noise and we used a high pass filter in second category to eliminated and motion artifact located in very low frequency band.
The remaining components from each category will be denoised via wavelet denoising with different levels. By the end, a nonlocal mean algorithm is used to enhance the results.

The rest of this paper is organized as follows. Section II describes the analysis and reduction of ECG signal noise using four approaches, low pass filter, Wavelet, and VMD.
In Section III, a complete description of the proposed methods is given. Each method aims to eliminate noise and artifacts.
Results and Discussions are in Section IV that show and compare the performance of the proposed methods via different simulations and using different metrics. Finally, Conclusions and recommendations are discussed in Section V.

## II. Analysis and elimination of ECG signal noise.

### 1) Analysis of ECG signal noise

Figure 1 shows that ECG signals have been distorted by a noise with quasi-harmonic behavior [16]. This observation is confirmed by its representation in the Frequency domain. The predominant energy of the noise is centered at a frequency of 50Hz.

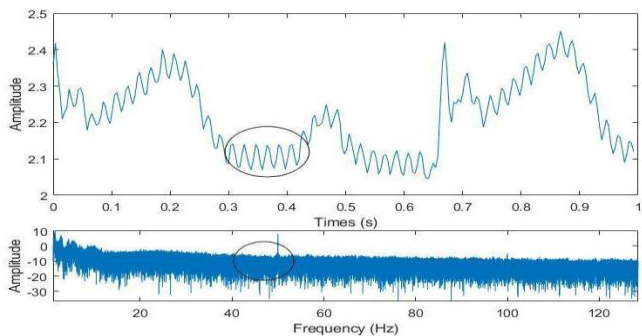

Fig. 1: An example of an ECG signal corrupted by noise in time and frequency domain.

There are several de-noising methods in the literature, including the low-pass filter, wavelet [17], EMD [18], VMD [11], LMS [19], and so on.
In order to find the most efficient way for effectively reducing noise while conserving signal information, we examine and compare three signal denoising methods: low-pass filter, wavelet, and VMD.

### 2) Signal denoising based on Low-Pass filter

Given the quasi-harmonic nature of the noise in the ECG signal, it is widely acknowledged that a low-pass filter with a cutoff frequency below Fc = 50 Hz, which has modest computational requirements, is useful in decreasing the noise.

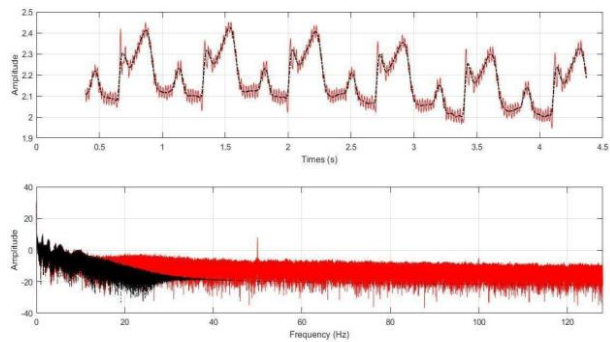

Fig. 2: Denoised ECG signal by using a low-pass filter with Fc = 15 Hz in time and frequency domain

We conducted several experiments and found that if the Fc=15Hz, the noise is reduced, as seen in Fig.2. However, various amplitudes of the signals are attenuated, indicating that some information in the signal is damaged, and therefore the quality of information is reduced. To address this issue, we increased the frequency cutting to 40Hz and found that less information is lost in the signal, although some noise occurs, as shown in Fig-3.

We conclude that it is difficult to achieve a good compromise between reducing noise while preserving signal information by using low-pass filters.

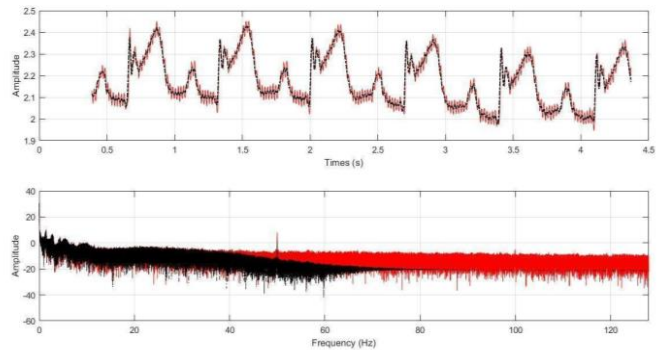

Fig. 3: Denoised ECG signal by using a low-pass filter with Fc = 40 Hz in time and frequency domain

### 3) Denoising signal based on Wavelet

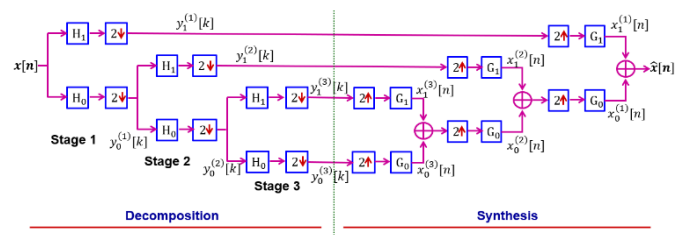

Fig. 4: Process of Decomposition and synthesis of signal by using discrete Wavelet [20].

The Wavelet decomposition technique, shown in Figure 4, involves the partitioning of a signal's frequency range into many sub-bands using two filters: a low pass filter and a high pass filter. At level 1, the signal is divided into two distinct bands: the high pass band and the low pass band. In the second level, the low pass band is divided into two independent bands: a low pass band and a high pass band. Each succeeding level involves splitting the low frequency band into two bands of equal size, as shown in Figure 4. The presence of down sampling after filtering ensures the transition from one level to the next [20]. As shown in Fig. 4, the sequences for level k are derived from those sequences by employing low-pass filters $h_0$ and $h_1$, respectively, as:

$$y_i^k(n) = \sum_l y_i^{k-1}(n) h_i(2l - n) \quad (1)$$

This decomposition approach effectively encircles and eliminates noise in a specific sub-band. As a result, the signal in this sub-band is disregarded throughout the reconstruction process to ensure that the final signal is free from any noise interference. The majority of the interference in electrocardiogram (ECG) signals is concentrated in the high-frequency sub-bands, namely around 50 Hz in our recorded data. With a frequency sample of 360 Hz, we need to use a decomposition level of at least 3 in order to remove noise. As a result, ⅞ of the frequency band, namely the high frequency band, will be discarded.

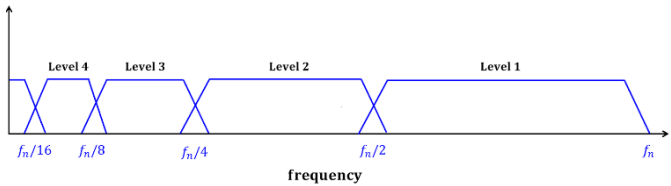

Fig-5. Frequency domain representation of the DWT

### 4) Signal denoising based on VMD

The most up to date signal processing technique is variational mode decomposition (VMD), which separates the input signal into several band-limited Intrinsic Mode Functions (IMFs).

The approach, inspired by the EMD [21], suggests that the original signal s(t) is composed of many IMFs known as FM components [22]. That is:

$$s(t) = \sum_k u_k(t) = \sum_k A_k(t) \cos(\phi_k(t)) \quad (2)$$

where $A_k(t)$ represents the instantaneous amplitude of $u_k(t)$ and $\phi_k(t)$ represents the instantaneous phase of $u_k(t)$. The IMF's center frequency $w_k(t)$ is presumed to be the correlating instantaneous frequency $w_k(t) = \phi'_k(t)$.

The decomposed VMDs, $u_k$, are compact around the frequency center $w_k(t)$. In VMD, $u_k$ and $w_k$ are calculated by solving the constrained variational problem as fellows:

$$\left\{ \sum_k \left\| \partial_t \left[ \left( \delta(t) + \frac{j}{\pi t} \right) * u_k(t) \right] e^{-jw_k t} \right\|_2^2 \right\} \quad (3)$$

The main advantage of this technique is that it is able to decompose the signal to different IMFs where each IMF has a specific frequency center and with a narrow band limited.

This decomposition has many benefits such as denoising signals and artifacts removal and it can be used for feature extraction.

We note that In Wavelet, the frequency band is divided into sub-bands where the size of each sub-band is related to a number of levels. From level to another, the size of the sub-band is divided by 2 (see Fig. 5).

Therefore, the width of each sub-band in Wavelet is fixed and is determined by the chosen level number. On the other hand, VMD decomposes signals into different IMFs. Each IMF has a specific frequency center and narrow sub-band. This offers more degrees of freedom to explore the frequency band of the signal.

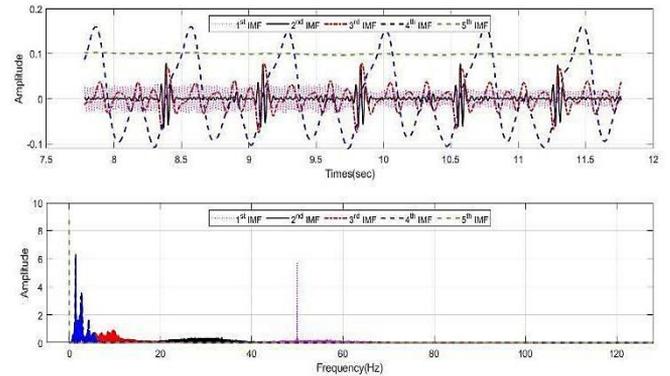

Fig-6. Representation of 5 IMFs of ECG signal in time and Frequency domains

Figure 6 clearly shows that each IMF has a distinct frequency center and a limited sub-band. Furthermore, harmonic noise appeared in the first IMF with frequency center of 50 Hz.

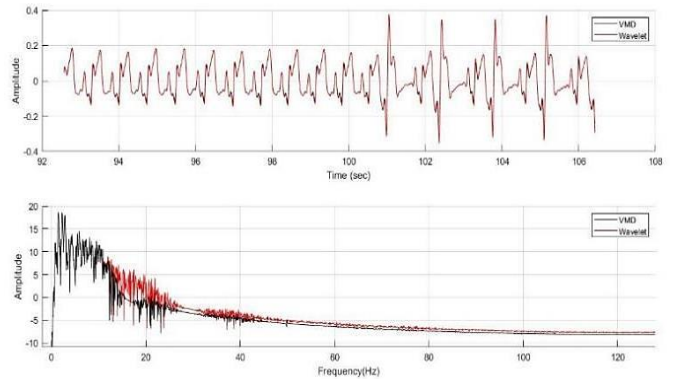

Fig-7. Representation of clean ECG signal in time and Frequency domains by using VMD and Wavelet Methods

We can observe that the denoising outcomes for the two approaches are almost similar. However, we see less interference between sub-bands in VMD compared to Wavelet. Due to the low computational complexity of Wavelet compared to VMD (adaptive algorithm that requires many iterations), Wavelet is more suitable for denoising ECG signals than VMD.

### III. Motion Artifacts Detection and Elimination in ECG signals

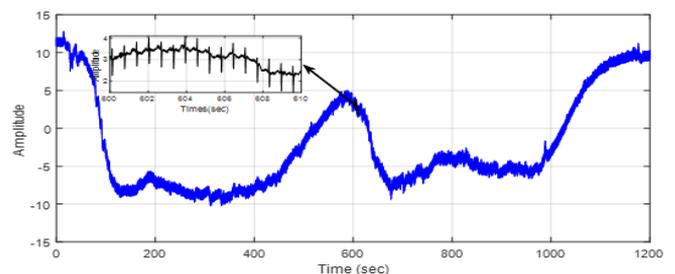

Fig-8. ECG signal with only motion artifact

By observing multiple ECG signal recording segments, we noted that in addition to the presence of noise, two types of motion artifacts can be observed. The first one acts on the ECG by moving segments of the ECG without deforming or changing the shape of the segments. This type of artifact can be easily observed in Figures 8 and 9.

Figure 9 demonstrates that some of the ECG signal segments are contaminated, and their forms are altered by the second kind of artifact. This artifact often manifests when movement is executed by a jump.

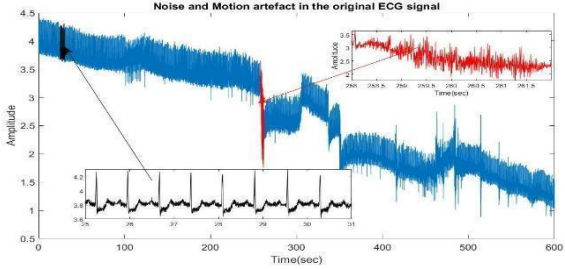

Fig-9. ECG signal contains two types of artifacts among them one is a non-stationary artifact signal.

### 1) Analysis of ECG signals corrupted by noise and artifacts

Based on the analysis of the ECG signal seen in Figure 8, we express this signal in the following form:
$$s(t) = s_c(t) + a(t) + n(t); \quad (4)$$
where $s_c(t)$ represents the clean signal without artifacts and noise, $n(t)$ and $a(t)$ represents the noise and artifact signal, respectively.

It was observed that in the frequency domain, the artifact signal is located in a very low-frequency band.
There are several methods for extracting this signal from the original ECG signals, including Wavelet, VMD and regression methods such as linear regression, robust linear regression, quadratic regression, Robust quadratic regression, etc.
Considering that Wavelet and VMD have previously been used for denoising purposes, we will further investigate them to effectively remove motion artifacts.
Before introducing the proposed methods for ECG signal denoising and artifacts eliminations.
We briefly describe Lilliefors Test and non-local means algorithms. Both techniques will be used in the proposed methods.

### 2) Lilliefors test

Lilliefors [23] presented a table for validating normality using the Kolmogorov-Smirnov statistic in cases where the mean and variance of the population are not known. His statistic is:
$$D_{max} = max_{\forall x} |S(x) - F^*(x)|, \quad (5)$$
where $S(x)$ is the sample distribution function and $F^*(x)$ is the cumulative normal distribution, whose mean and variance are determined by the sample.
The null hypothesis that the sample is normally distributed is rejected if the computed statistic test $D_{max}$ exceeds the critical value for the chosen significance level.

### 3) Nonlocal means algorithm:

Non-local means (NLM) denoising for signals entails calculating weighted averages of signal samples based on their similarity to other signals.
$$Let\ x(n) = u(n) + v(n)$$
where, $u(n), and\ v(n)$ represent the clean signal and the noise, respectively. Any sample $u(m)$, can be estimated as follows

$$\hat{u}(m) = \frac{1}{T(m)} \sum_{n \in N(m)} w(m,n)\, x(n) \quad (6)$$

where, $w(m,n)$ is the weight associated with $n-th$ search sample and $m-th$ desired sample in a search window $N(m)$ and $T(m) = \sum_n w(m,n)$. The weight is given by:
$$w(m,n) = exp\left(\frac{\sum_{\Delta \in \delta}(x(m+\Delta) - x(n+\Delta))^2}{2P_\delta k^2}\right) \quad (7)$$

where, $k, \delta$ are the bandwidth parameter and local patch of sample surrounding the $m-th$ desired sample containing $P_\delta\ samples$.

### 1) Noise and Motion artifact removal Based on Wavelet, Lilliefors test, NLM and high-pass Filter (WLNH method)

We presented a novel method based on Wavelet multi-resolution analysis (MRA) to simultaneously eliminate motion artifacts and noise. Afterwards, the ECG signal is broken down into its component parts. There is a set bandwidth for each component.

According to Figure 10, we set the number of wavelet levels to 10. Consequently, we acquired a total of 11 components: the first 10 components are referred to as details, while the last component represents the estimated signal at level 10.

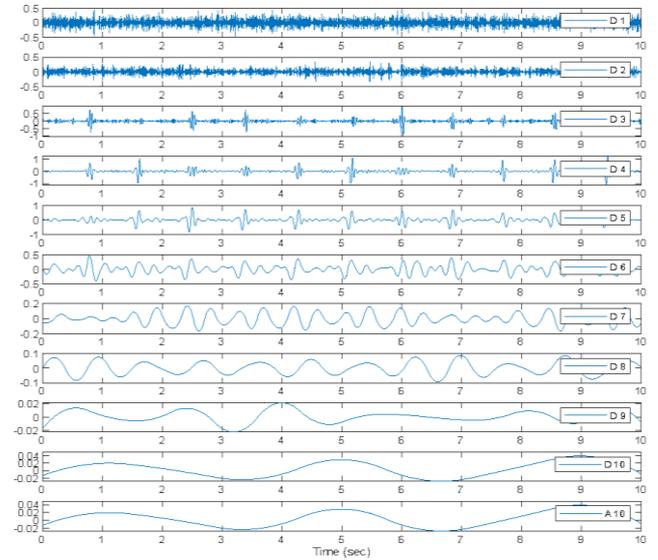

Fig-10. An example of MRA Wavelet at level 10 of ECG Signal

We categorize the components into two distinct groups. The first category comprises the initial 5 components that are tainted by noise. As seen in Figure 11, these components are in high frequency ranges, and they cover more than 95% of the frequency spectrum (precisely 96.88%). The second group has components numbered 6 to 11. This group is distinguished by the scarcity of sounds. We also see the occurrence of motion artifacts in the final components in several instances.

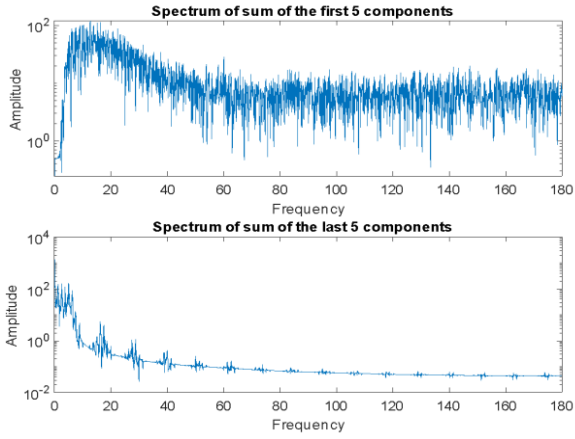

Fig. 11 Spectrum of the first 5 signal details and the spectrum of the sum of last 5 signal details with signal approximation at level 10.

Based on aforementioned information, we propose the following 6-stages algorithm, whose block diagram is shown in Fig. 12.

**Proposed Method (WLNH Method):**
Stage 1: Decompose the ECG signal into 11 components using multiresolution Discrete Wavelet (10 levels).
Stage 2: Apply the Lilliefors test [23] to each of the first five components. The Lilliefors test determines whether the noise is white Gaussian or not. If the test results are positive for any component, that component will be deleted.
Stage 3: The remaining components from the five components will be combined to create a signal. We utilize a DWT to denoise this signal at a somewhat higher level.
Stage 4: As seen in figure 12, the total of the final six components includes a little amount of noise. To denoise this signal, we employ DWT at low levels.
Stage 5: The total of the noised 5 components is added to the sum of the last denoised components, and the resulting signal is filtered via a high pass filter to eliminate motion artifacts.
Stage 6: Finally, we apply the NLM algorithm to the filtered signal to increase the quality of the denoised signal while removing artifacts.

The flowchart in Fig. 12 depicts the many processes employed in the suggested technique.

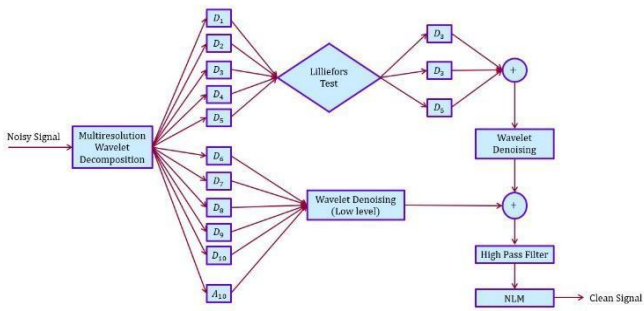

Figure 12 Flowchart of the proposed method 1 (WLHN Method)

**2) Noise and Motion artifact removal Based on VMD Lilliefors test, Wavelet, High Pass Filter and NLM (VLWNH)**

Variational Mode decomposition offers a new alternative to decompose the signal in many components called IMfs where each IMF has a specific central frequency with a narrow bandwidth.

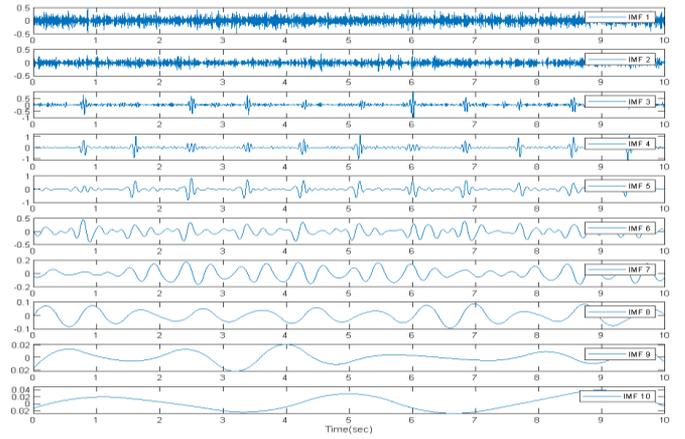

Figure 13: An example *VMD decomposition result for ECG signal when the mode is 10*

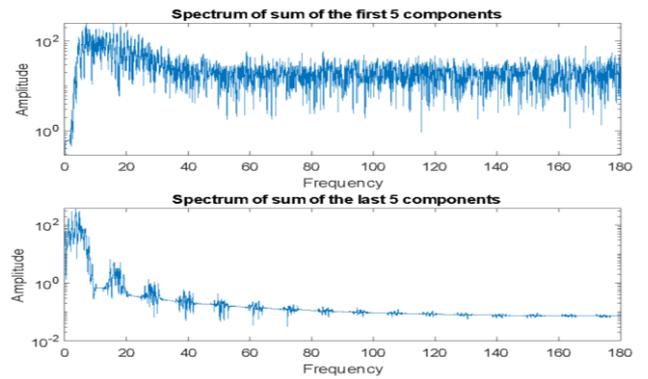

Fig14. Spectrum of the sum of the first 5 IMFs and the spectrum of the sum of the last 5 IMFs.

Figures 13 and 14 illustrate that VMD and MRA Wavelet decompositions have roughly identical components, as well as signal partitioning in the frequency domain. It is obvious that the sum of the first five IMFs filled the biggest signal spectrum range, but the final five IMFs occupied a relatively tight low frequency band.

Furthermore, we note that most of the noise is concentrated in the initial IMFs.

We then presented a novel method, similar to the one described in the preceding section (section III. 3), by simply substituting MRA Wavelet with VMD. The following flowchart depicts the several processes in this Method.

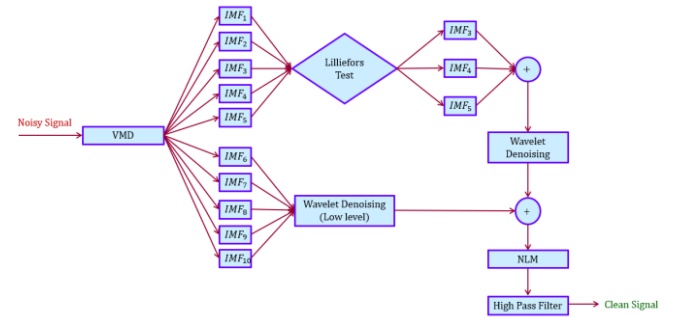

Figure 15 Flowchart of the proposed method 2 (VWHN Method)

## IV. Results and Discussion

**1) Test databases:**
To compare the performance of the proposed approaches to those in the literature, we used three reference signals from the MIT-BIH Arrhythmia database. The signals are 100, 103, and 105. Each recording lasts 30 minutes and uses a sample rate of 360 Hz. We also employed three kinds of noise. The first is white Gaussian noise (WGN), while the others are base-line wander (BW) and muscle artifacts (MA) from the MIT-BIH Noise Stress Test Database.

### 2) Performance Metrics
To compare various denoising approaches, four standard metrics are employed.

a) Mean Square Error (MSE) is calculated as the average of the squared differences between the denoised ECG signal $\hat{x}(n)$ with size N and the original clean ECG signal $x(n)$ with the same size. The MSE is given by
$$MSE = \frac{1}{N}\sum_{n=0}^{N-1}(x(n) - \hat{x}(n))^2 \quad (8)$$

b) Root Mean Square Error (RMSE) is the square root of MSE defined as:
$$RMSE = \sqrt{MSE} \quad (9)$$
The approach with the lowest MSE/RMSE value achieves the highest performance.

c) Percentage-Root-mean-square Difference (PRD) calculates the percentage of overall distortion in the signal after denoising. A lower PRD indicates a denoised signal with higher quality.
$$PRD = \sqrt{\frac{\sum_{n=0}^{N-1}(\hat{x}(n)-x(n))^2}{\sum_{n=0}^{N-1}(x(n))^2}} \times 100\% \quad (10)$$

d) Improved signal-to-noise ratio ($SNR_{imp}$)

The term "improved signal-to-noise ratio" (SNR_imp) refers to an improvement in signal quality when noise reduction methods are employed. It measures how much better the signal is than the noise after processing.

$$SNR_{imp} = SNR_{out} - SNR_{in}$$
Where
$$SNR_{in} = 10 \times \log_{10}\left(\sqrt{\frac{\sum_{n=0}^{N-1}(x(n))^2}{\sum_{n=0}^{N-1}(\hat{x}(n)-x(n))^2}}\right) \quad (11)$$
$$SNR_{out} = 10 \times \log_{10}\left(\sqrt{\frac{\sum_{n=0}^{N-1}(x(n))^2}{\sum_{n=0}^{N-1}(\hat{x}(n)-x(n))^2}}\right) \quad (12)$$

### 3) PERFORMANCE EVALUATION OF THE PROPOSED METHODs

To demonstrate the effectiveness of our proposed methods, we analyzed their performance under various noise and artifact settings.
We used three forms of noise: additive white Gaussian noise (AWGN), base-line wander (BW), muscle artifacts and (MA).

To ensure accurate results, the proposed methods are evaluated for each 10-second ECG segment, which consists of 3600 samples. The final results are calculated by taking the average of the results from 180 segments, with a total duration of 30 minutes.

The Lilliefors test [23] is employed in each ECG segment in order to determine whether a component is white Gaussian noise or not. The test is conducted with a significance level of 5%. We employ an infinite impulse response (IIR) filter and a high pass filter with a frequency reduction of 3 Hz, as outlined in the proposed methods.

In Wavelet MRA, we used 10 levels. The wavelet mother "Fk14 " is used in both Wavelet MRA decomposition and also in wavelet denoising algorithm. Different levels of wavelet denoising are used in the first 5 components and in the last components.

#### a) Performance Evaluation in the presence of AWGN on MIT-BIH databases
In the presence of white Gaussian noise, it was revealed that the first five wavelet MRA components (also known as the first five IMFs) have at least one Gaussian white noise component.
We demonstrate that both the wavelet MRA and VMD are capable of separating some white Gaussian noise from the noisy data. Figure 20 clearly shows that the probability plot of the first component corresponds perfectly to the typical probability plot.
Using the Lilliefors test, any component indicating white Gaussian noise is suppressed, resulting in reduced noise quality in the noisy ECG signal.

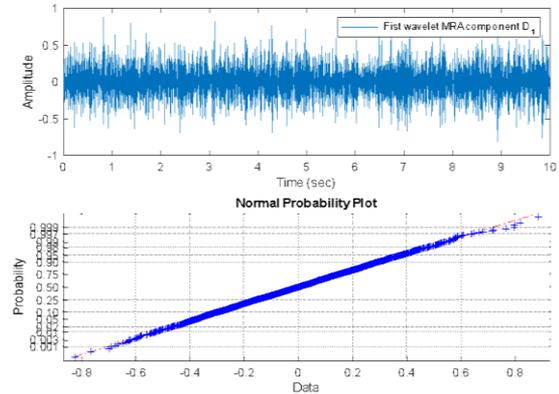

*Figure 20: First Wavelet MRA component and its Normal Probability plot*

This research used the MIT-BIH arrhythmia database's dataset of ECG record 100 for qualitative analysis. The addition of white noise to an existing ECG signal makes it loud. The noisy ECG signal's signal-to-noise ratio (SNR) remains at 0 dB. Fig. 21 indicates that the proposed method successfully denoised the ECG signal, which retains crucial morphological information from the original signal and achieves a high degree of similarity to the clean ECG signal.
The proposed methods in this paper can be looked at as an important extension of the proposed method in [13].
In order to assess the quantitative performance of the proposed methods, we used table 5 in [13] to compare our findings with other current methods such as VMD-NLM [13], EMD-wavelet [24], NLM-MEMD [25], and NLM-DWT [26]. We limit benchmarking to ECG recording MIT-BIH arrhythmia database 100, 103,105.

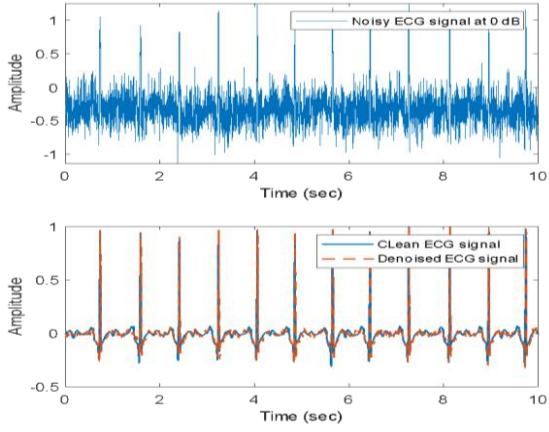

Figure 21 A closer comparison of denoised and original clean ECG signal (dashed line). The noisy ECG signal at 0 dB input SNR is also plotted.

| ECG | Measure | WLNH | VLWNH | VMD-NLM | EMD-Wavelet | NLM-MEMD | NLM-DWT |
|---|---|---|---|---|---|---|---|
| 100 | MSE$\times 10^{-4}$ | 2.3 | 2.5 | 3.0 | 9 | 8.0 | 3.9 |
| | PRD % | 4.67 | 4.37 | 5.28 | 9.23 | 8.11 | 5.5 |
| | SNRimp | 26.59 | 26.12 | 8.92 | 7.34 | 5.21 | 8.58 |
| 103 | MSE$\times 10^{-4}$ | 1.8 | 1.8 | 8.0 | 25 | 1.7 | 8.9 |
| | PRD % | 5.02 | 5.04 | 7.63 | 13.43 | 10.89 | 7.75 |
| | SNRimp | 27.29 | 27.4 | 8.56 | 7.64 | 7.54 | 8.58 |
| 105 | MSE$\times 10^{-4}$ | 7.2 | 7.1 | 18.0 | 22 | 20.0 | 11.0 |
| | PRD % | 5.76 | 6.18 | 8.69 | 8.87 | 12.05 | 8.87 |
| | SNRimp | 22.57 | 22.64 | 8.3 | 8.16 | 5.51 | 8.16 |

Table 1: Performance of the proposed methods and explored methods on the test dataset ECG 100 at input SNR level of 10 dB

Table 1 shows the performance of the denoising algorithms on the test datasets ECG 100, 103, and 105 at an input SNR level of 10 dB in terms of MSE, PRD, and SNRimp. We definitely observed that the proposed strategies outperformed current methods for all three records in terms of MSE, PRD, and SNRimp. As example, the MSE of proposed methods in record 105 is $7.1 \times 10^{-4}$ and the best method in the litterature has $11 \times 10^{-4}$

All recordings show a significant change, particularly records 103 and 105.
We also observe that the performance of the two suggested methods, WLNH and VLWNH, are almost identical, with minimal benefit to VLWNH.

### b) Performance Evaluation in the presence of BW noise on MIT-BIH databases

We evaluate the performance of the proposed methods in the presence of base-line wander (BW) at two distinct SNR levels of 0 and 5 dB. We compare the performance of the suggested strategy to other methods described in table 10 in [10].
As demonstrated in Table 2, the proposed strategies outperform the current conventional methods including Wavelet Transform (WT) [27] and also for some approach based on deep learning as stacked DAE [28], improve DAE [29]. The only approach based on deep learning GAN [30] outperforms the offered techniques, although our findings are comparable. We also emphasize that every deep learning approach has a high computational cost.

Thus, we may conclude that the offered approaches provide the optimal trade-off between performance and computational complexity. We also observe that the suggested methods' performance is pretty similar to each other. So, for the next cases, we will only evaluate the WNH approach due to its lower complexity compared to VWNH.

| ECG | SNR | Performance Measure | WNH | VWNH | GAN | stacked DAE | Improved DAE | WT |
|---|---|---|---|---|---|---|---|---|
| 103 | 0dB | RMSE$\times 10^{-3}$ | 6 | 5.9 | 3.2 | 38 | 26 | 74 |
| | | PRD | 1.31 | 1.39 | 0.97 | 9.75 | 6.47 | 18.05 |
| | | SNRimp | 37.41 | 37.31 | 40.26 | 20.38 | 23.78 | 14.87 |
| | 5dB | RMSE$\times 10^{-3}$ | 4.5 | 4.4 | 2.7 | 37 | 25 | 74 |
| | | PRD | 1.13 | 1.09 | 0.83 | 9.15 | 6.39 | 17.99 |
| | | SNRimp | 37.75 | 37.89 | 41.60 | 15.77 | 18.89 | 9.9 |
| 105 | 0dB | RMSE$\times 10^{-3}$ | 12.6 | 1.33 | 3.5 | 29 | 28 | 14 |
| | | PRD | 2.3 | 2.42 | 1.06 | 5.69 | 5.37 | 2.65 |
| | | SNRimp | 32.1 | 31.44 | 39.49 | 24.9 | 25.4 | 31.53 |
| | 5dB | RMSE$\times 10^{-3}$ | 9.5 | 9.6 | 3.4 | 27 | 27 | 12 |
| | | PRD | 1.5768 | 1.51 | 0.094 | 5.33 | 5.34 | 2.31 |
| | | SNRimp | 32.77 | 32.49 | 40.56 | 20.47 | 20.45 | 27.71 |

Table 2: Performance of the proposed methods and explored methods on the test dataset ECG 103 and 105 at input SNR level 0 dB and 5dB of base-line wander (BW) noise.

### c) Performance of various methodologies in the presence of MA on MIT-BIH databases

Table 3 compares the performance of several approaches for ECG signal denoising and artifact elimination when muscular artifacts (MA) noise is present at SNR values of 0dB and 5dB. The GAN [30] approach, based on deep learning, still produces the best results. Except for the GAN approach, all methods perform similarly owing to the non-stationary partitioning of noise in the ECG signal. In reality, certain ECG segments are much noisier than the rest of the signal.

| ECG Record | SNR | Perf. Measure | WLNH | GA | stacked DAE | DAE | WT |
|---|---|---|---|---|---|---|---|
| 103 | 0dB | RMSE | 0.022 | 0.004 | 0.046 | 0.034 | 0.044 |
| | | PRD | 6.36 | 0.86 | 11.32 | 8.53 | 10.4 |
| | | SNRimp | 25.77 | 41.36 | 18.92 | 21.38 | 19.66 |
| | 5dB | RMSE | 0.016 | 0.003 | 0.044 | 0.027 | 0.067 |
| | | PRD | 3.98 | 0.69 | 10.83 | 6.82 | 16.24 |
| | | SNRimp | 27.16 | 38.24 | 14.31 | 18.33 | 10.79 |
| 105 | 0dB | RMSE | 0.0436 | 0.007 | 0.036 | 0.03 | 0.04 |
| | | PRD | 8.84 | 1.5 | 7.1 | 5.81 | 7.86 |
| | | SNRimp | 21.6219 | 36.49 | 22.97 | 24.72 | 22.09 |
| | 5dB | RMSE | 0.031 | 0.005 | 0.032 | 0.028 | 0.032 |
| | | PRD | 7.4772 | 1.05 | 6.22 | 5.54 | 6.23 |

| | SNRimp | 22.3689 | 34.55 | 19.12 | 20.13 | 19.11 |

Table 3 : Performance of the proposed methods and explored methods on the test dataset ECG 103 and 105 at input SNR level 0 dB and 5dB of by muscle artefacts (MA) noise

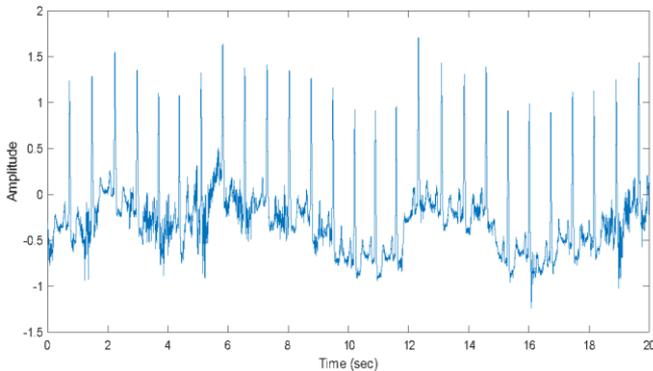

Fig. 22 An example of Noisy ECG record 105 corrupted by muscle artefacts (MA) noise

Fig .22 illustrates the variation of the noise from segment to another. These situations open the door for the possibility of new improvement of the proposed algorithms in order to take into account the variation of noise in short duration.

## V. Conclusion

In this study, we presented two novel methods for denoising and artifact removal in ECG signals.

The proposed methods incorporate a variety of signal processing and statistical techniques. In fact, we used Wavelet MRA/VMD for signal decomposition, Lillfores test, high pass filter, wavelet denoising, and the nonlocal mean algorithm to improve the performance of the proposed approaches.

The performances of the new methods are illustrated and compared to existing ones. In reality, the proposed approaches perform well when the noise is additive white gaussian noise, as well as when there are base-line wander artifacts. In However, due to the presence of noise with varying levels in the same segment of ECG signals (for example, MA noise), the performance of the suggested approaches can be enhanced to take this into account. A new challenge has been established to improve the proposed methods by considering this context.